\documentclass[runningheads]{llncs}
\usepackage[T1]{fontenc}
\usepackage{graphicx}
\usepackage{booktabs}
\usepackage{hyperref}
\usepackage[misc]{ifsym}

\usepackage{mwe}
\usepackage{bm}
\usepackage{amsfonts}
\usepackage{amsmath}
\usepackage{tikz}
\usepackage{comment}

\begin{document}

\newcommand{\manu}[1]{\textcolor{red}{[#1]}}

\title{Staged Event Trees for Transparent Treatment Effect Estimation}

\titlerunning{Staged Event Trees for Transparent Treatment Effect Estimation}

\author{Gherardo Varando\inst{1} \and
Manuele Leonelli \inst{2} \and
 Jordi Cerdà-Bautista \inst{3} \and 
 Vasileios Sitokonstantinou \inst{4} \and 
 Gustau Camps-Valls \inst{3}}

\authorrunning{G. Varando et al.}

\institute{Department of Statistics and Operational Research, University of Valencia, Valencia, Spain  \email{gherardo.varando@uv.es}
\and
School of Science and Technology, IE University, Madrid, Spain \email{manuele.leonelli@ie.edu}
\and
Image Processing Laboratory, University of Valencia, Valencia, Spain \email{\{jordi.cerda,gustau.camps\}@uv.es}\and
Artificial Intelligence Group, Wageningen University \& Research, Wageningen, The Netherlands \email{vasileios.sitokonstantinou@wur.nl}
}

\maketitle              

\begin{abstract}
  Average and conditional treatment effects are fundamental causal quantities used to evaluate the effectiveness of treatments in various critical applications, including clinical settings and policy-making. Beyond the gold-standard estimators from randomized trials, numerous methods have been proposed to estimate treatment effects using observational data. 
  In this paper, we provide a novel characterization of widely used causal inference techniques within the framework of staged event trees, demonstrating their capacity to enhance treatment effect estimation.
  These models offer a distinct advantage due to their interpretability, making them particularly valuable for practical applications. We implement classical estimators within the framework of staged event trees and illustrate their capabilities through both simulation studies and real-world applications. Furthermore, we showcase how staged event trees explicitly and visually describe when standard causal assumptions, such as positivity, hold, further enhancing their practical utility.
\keywords{Staged event tree  \and average treatment effect \and causal inference.}
\end{abstract}

\section{Introduction}

Causal inference seeks to estimate causal effects from data by combining statistical models, assumptions, and observed information~\cite{hernancausal,pearl2009causal,pearl2016causal,peters2017elements}. 
The primary and most used target of interest is the causal treatment effect, also called the average treatment effect (ATE), which quantifies the change in the expected value of the outcome of interest under 
different interventions or treatment values~\cite{hernancausal}. 
Randomized controlled trials (RCTs) are considered the gold standard for estimating 
ATEs because randomization balances both observed and unobserved confounders~\cite{hernancausal}. 
However, RCTs are often impractical due to ethical, logistical, or financial constraints. This limitation requires estimating causal effects from observational data, prompting the development of various causal inference methodologies~\cite{hernancausal,runge2023causal}.

Probabilistic graphical models, especially directed acyclic graphs (DAGs), have been widely used in causal analysis. DAGs provide an intuitive and practical framework for representing prior knowledge and assumptions. They enable causal reasoning processes to determine the identifiability of causal quantities and derive non-parametric estimators.~\cite{huang2006identifiability,pearl2009causal,shpitser2006identification}. 

Despite their widespread use, DAGs have limitations in representing context-specific independencies and asymmetric relationships inherent in complex systems. To address these challenges, \emph{staged event trees} have emerged as a flexible class of probabilistic graphical models grounded in event trees~\cite{collazo2018chain,smith2008conditional}. Staged event trees extend DAGs by graphically encoding \emph{asymmetric} conditional independence statements, such as context-specific independencies~\cite{pensar2016role}, allowing for more nuanced patterns of dependence in categorical data.

Recent advances have enriched the theoretical foundations of staged event trees~\cite{varando2024staged} and led to the development of efficient algorithms for their construction and analysis~\cite{leonelli22a,leonelli2024structural}. These advances, combined with software implementations, have greatly facilitated the adoption of staged event trees in practical applications~\cite{carli2022,walley2023cegpy}. Staged event trees have already proven valuable in the realms of causal discovery and analysis~\cite{cowell2014causal,leonelli23a,thwaites2010causal}. Notably, recent studies highlight that incorporating context-specific information - visually captured by staged event trees - significantly enhances the identification of causal effects, suggesting a potential practical impact of staged event trees in complex causal inference tasks~\cite{mokhtarian2022causal,tikka2019identifying}.


This study addresses two pivotal aspects of causal inference. First, we introduce a novel integration of causal inference techniques into the staged event tree framework, broadening their capability to analyze complex causal relationships. Second, we demonstrate the practical insights that staged event trees offer in applied causal analyses, emphasizing their strength in visualizing and estimating causal effects in a clear and interpretable way. 
With their inherent flexibility and transparency, staged event trees have the potential to reshape how applied causal analyses are conducted, offering a more intuitive and structured approach to estimating and visualizing causal effects in real-world settings.


The paper is structured as follows. After reviewing the theory of staged event trees in Section~\ref{sec:SET}, we show how the ATE and its conditional version (CATE) can be easily computed from fitted staged event trees (Section~\ref{sec:causalinference}). We showcase the performance of the derived estimators in a simulation study (Section~\ref{sec:experiments}) and classical datasets from the causal inference literature (Section~\ref{sec:experiments2}). We conclude in Section~\ref{sec:conclusions} with final remarks and an outlook.

\section{Staged Event Trees}\label{sec:SET}

Staged event trees are a class of probabilistic graphical models that extend traditional event trees by incorporating context-specific conditional independence structures. They provide a flexible and interpretable framework for modeling complex relationships among categorical variables \cite{collazo2018chain,smith2008conditional}.

Let $[p]=\{1,\dots,p\}$ and $\bm{X}=(X_1, X_2, \dots, X_p)=(X_i)_{i\in[p]}$ be a sequence of categorical random variables with joint probability mass function $P$ and sample space $\mathbb{X}=\prod_{i=1}^p \mathbb{X}_i$, where each $\mathbb{X}_i$ is the finite set of possible values of $X_i$. For any subset $A\subseteq [p]$, let $\bm{X}_A=(X_i)_{i\in A}$ and $\bm{x}_A=(x_i)_{i\in A}\in\mathbb{X}_A=\prod_{i\in A}\mathbb{X}_i$. We also write $\bm{X}_{-A}=\bm{X}_{[p]\setminus A}$.


Let $(V,E)$ be a finite, rooted, directed tree with vertex set $V$, edge set $E$, and root node $v_0$. For each $v\in V$, let $E(v)=\{(v,w)\in E\}$ be the set of edges emanating from $v$. Let $\mathcal{C}$ be a finite set of labels.

\begin{definition}
\label{def:x}
An $\bf X$-compatible staged event tree 
is a triplet $T = (V,E,\eta)$, where $(V,E)$ is a rooted directed tree and
\begin{enumerate}
    \item[i)] $V = {v_0} \cup \bigcup_{i \in [p]} \mathbb{X}_{[i]}$;
		\item[ii)] For all $v,w\in V$,
$(v,w)\in E$ if and only if $w=\bm{x}_{[i]}\in\mathbb{X}_{[i]}$ and 
			$v = \bm{x}_{[i-1]}$, or $v=v_0$ and $w=x_1$ for some
$x_1\in\mathbb{X}_1$;
\item[iii)] $\eta:E\rightarrow \mathcal{L}=\mathcal{C}\times \cup_{i\in[p]}\mathbb{X}_i$ is a labelling of the edges such that $\eta(v,\bm{x}_{[i]}) = (\kappa(v), x_i)$ for some 
			function $\kappa: V \to \mathcal{C}$. 
\end{enumerate}
If $\eta(E(v)) = \eta(E(w))$ then $v$ and $w$ are said to be in the same \emph{stage} in $T$.
\end{definition} 
Therefore, the equivalence classes induced by  $\eta(E(v))$ form a partition of the internal vertices of the tree in \emph{stages}. 

Points \emph{i}) and \emph{ii}) in Definition~\ref{def:x} define a classical event tree for the random vector $\mathbf{X}$. In particular, internal nodes correspond to the elements of the sample spaces $\mathbb{X}_{[i]}$ and leaves correspond to the elements in $\mathbb{X}$. 
Nodes are then connected by edges, following the order of the variables in $\mathbf{X}$, in such a way that a classical event tree is constructed.
Point \emph{iii}) introduces a labeling of edges that is compatible with the sample spaces of the respective random variables. Such labeling $\eta$ can be better represented graphically by the \emph{coloring} $\kappa$ of the internal nodes, and it induces the stages partition.  

Every $\bm X$-compatible staged event trees is linked to  a statistical model by associating conditional probabilities to the event tree in the standard way and, additionally, imposing equality of conditional probabilities respecting the \emph{stages structure} defined by $\eta$.
More formally, the \emph{staged event tree model} $\mathcal{M}_T$ is comprised of all 
joint probability for $\bm X$ such that, if we consider the chain factorization in the given order $P(\bm X) = P(X_1)P(X_2|X_1)\cdots P(X_p|\bm{X}_{[p-1]})$, conditional probabilities are equal when they correspond to nodes in the same stage in the tree $T$; that is, 
\[
P(X_i| \bm{X}_{[i-1]} = \bm{x}_{[i-1]}) = P(X_i| \bm{X}_{[i-1]} = \bm{x}'_{[i-1]}),
\]
if nodes  $\bm{x}_{[i-1]}$ and $\bm{x}'_{[i-1]}$  are in the same stage.



As detailed by~\cite{leonelli23a}, staged event trees can also be interpreted causally by defining an interventional causal model as a collection of interventional distributions in an intuitive way: the joint distribution of $\bf{X}$ in the intervened model is obtained by the product of the parameters along the corresponding root-to-leaf path, where (conditional) distributions for the intervened variables are replaced.

\begin{definition}[Staged tree causal model \cite{leonelli23a}]

A staged tree causal model induced by an
$\bm{X}$-compatible staged event tree $T = (V, E, \eta)$ is the finite class of  interventional distributions that are defined for each set of intervened variables $\bm{X}_I$ and their values:
  \begin{equation}
    \begin{split}
    &P\left(\bm{X} = \bm{x} | \operatorname{do}( \bm{X}_I = \bm{z}_I) \right) = \\ &
    \left\{ 
    \begin{matrix} 
    \frac{ P( \bm{X} = \bm{x} ) }{\prod_{i \in I} 
    P(X_i = x_i | \bm{X}_{[i-1]} = \bm{x}_{[i-1]}) }  & \text{if } \bm{x}_I = \bm{z}_I\\ 
    0   & \text{otherwise}
    \end{matrix} \right.
    \end{split}
    \end{equation}

\end{definition}

Staged event trees, when the number of variables remains moderate, offer the advantage of being easily visualized and understood by most practitioners, even those without a statistical background, due to their familiarity with event trees~\cite{filigheddu2024using}. This intuitive visual representation extends beyond simple event trees by allowing for a more flexible depiction of asymmetric conditional independence statements, while maintaining high interpretability. This flexibility proves especially valuable in analyzing causal effects and modeling possible interventions.

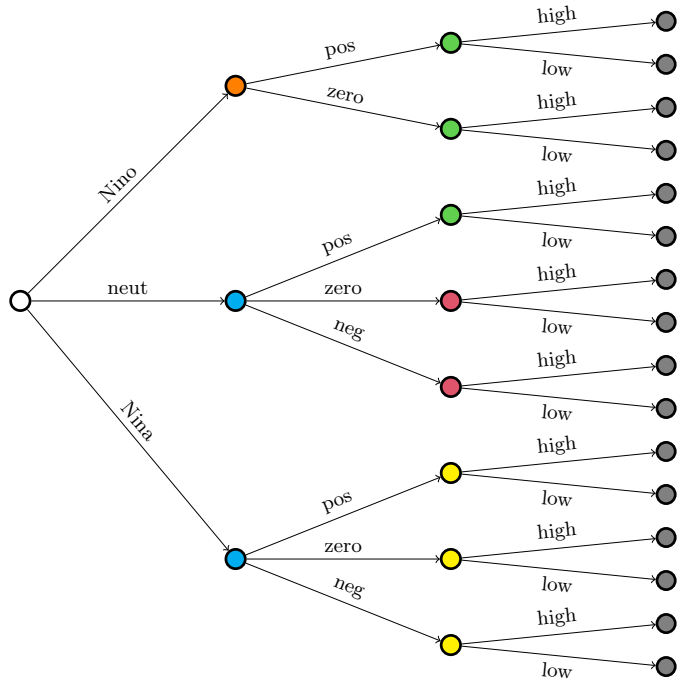
\begin{figure}[t!]
    \centering
    \resizebox{3\columnwidth/4}{!}{
    \begin{tikzpicture}[auto, scale=10,
	enso_NA/.style={circle,inner sep=1mm,minimum size=0.3cm,draw,black,very thick,fill=white,text=black},
	iod_2/.style={circle,inner sep=1mm,minimum size=0.3cm,draw,black,very thick,fill=cyan,text=black},
	iod_3/.style={circle,inner sep=1mm,minimum size=0.3cm,draw,black,very thick,fill=orange,text=black},
	au_3/.style={circle,inner sep=1mm,minimum size=0.3cm,draw,black,very thick,fill=yellow,text=black},
	au_5/.style={circle,inner sep=1mm,minimum size=0.3cm,draw,black,very thick,fill={rgb,255:red,223; green,83; blue,107},text=black},
	au_8/.style={circle,inner sep=1mm,minimum size=0.3cm,draw,black,very thick,fill={rgb,255:red,97; green,208; blue,79},text=black},
	au_UNOBSERVED/.style={circle,inner sep=1mm,minimum size=0.3cm,draw,black,very thick,fill={rgb,255:red,255; green,255; blue,255},text=black},
	 leaf/.style={circle,inner sep=1mm,minimum size=0.2cm,draw,very thick,black,fill=gray,text=black}]

	\node [enso_NA] (root) at (0.000000, 0.566667)	{};
	\node [iod_2] (root-Nina) at (0.333333, 0.166667)	{};
	\node [iod_2] (root-neut) at (0.333333, 0.566667)	{};
	\node [iod_3] (root-Nino) at (0.333333, 0.900000)	{};
	\node [au_3] (root-Nina-neg) at (0.666667, 0.033333)	{};
	\node [au_3] (root-Nina-zero) at (0.666667, 0.166667)	{};
	\node [au_3] (root-Nina-pos) at (0.666667, 0.300000)	{};
	\node [au_5] (root-neut-neg) at (0.666667, 0.433333)	{};
	\node [au_5] (root-neut-zero) at (0.666667, 0.566667)	{};
	\node [au_8] (root-neut-pos) at (0.666667, 0.700000)	{};
	\node [au_8] (root-Nino-zero) at (0.666667, 0.833333)	{};
	\node [au_8] (root-Nino-pos) at (0.666667, 0.966667)	{};
	\node [leaf] (root-Nina-neg-low) at (1.000000, 0.000000)	{};
	\node [leaf] (root-Nina-neg-high) at (1.000000, 0.066667)	{};
	\node [leaf] (root-Nina-zero-low) at (1.000000, 0.133333)	{};
	\node [leaf] (root-Nina-zero-high) at (1.000000, 0.200000)	{};
	\node [leaf] (root-Nina-pos-low) at (1.000000, 0.266667)	{};
	\node [leaf] (root-Nina-pos-high) at (1.000000, 0.333333)	{};
	\node [leaf] (root-neut-neg-low) at (1.000000, 0.400000)	{};
	\node [leaf] (root-neut-neg-high) at (1.000000, 0.466667)	{};
	\node [leaf] (root-neut-zero-low) at (1.000000, 0.533333)	{};
	\node [leaf] (root-neut-zero-high) at (1.000000, 0.600000)	{};
	\node [leaf] (root-neut-pos-low) at (1.000000, 0.666667)	{};
	\node [leaf] (root-neut-pos-high) at (1.000000, 0.733333)	{};
	\node [leaf] (root-Nino-zero-low) at (1.000000, 0.800000)	{};
	\node [leaf] (root-Nino-zero-high) at (1.000000, 0.866667)	{};
	\node [leaf] (root-Nino-pos-low) at (1.000000, 0.933333)	{};
	\node [leaf] (root-Nino-pos-high) at (1.000000, 1.000000)	{};

	\draw[->] (root) -- node [sloped]{Nina} (root-Nina);
	\draw[->] (root) -- node [sloped]{neut} (root-neut);
	\draw[->] (root) -- node [sloped]{Nino} (root-Nino);
	\draw[->] (root-Nina) -- node [sloped]{neg} (root-Nina-neg);
	\draw[->] (root-Nina) -- node [sloped]{zero} (root-Nina-zero);
	\draw[->] (root-Nina) -- node [sloped]{pos} (root-Nina-pos);
	\draw[->] (root-neut) -- node [sloped]{neg} (root-neut-neg);
	\draw[->] (root-neut) -- node [sloped]{zero} (root-neut-zero);
	\draw[->] (root-neut) -- node [sloped]{pos} (root-neut-pos);
	\draw[->] (root-Nino) -- node [sloped]{zero} (root-Nino-zero);
	\draw[->] (root-Nino) -- node [sloped]{pos} (root-Nino-pos);
	\draw[->] (root-Nina-neg) -- node [swap,sloped]{low} (root-Nina-neg-low);
	\draw[->] (root-Nina-neg) -- node [sloped]{high} (root-Nina-neg-high);
	\draw[->] (root-Nina-zero) -- node [swap,sloped]{low} (root-Nina-zero-low);
	\draw[->] (root-Nina-zero) -- node [sloped]{high} (root-Nina-zero-high);
	\draw[->] (root-Nina-pos) -- node [swap,sloped]{low} (root-Nina-pos-low);
	\draw[->] (root-Nina-pos) -- node [sloped]{high} (root-Nina-pos-high);
	\draw[->] (root-neut-neg) -- node [swap,sloped]{low} (root-neut-neg-low);
	\draw[->] (root-neut-neg) -- node [sloped]{high} (root-neut-neg-high);
	\draw[->] (root-neut-zero) -- node [swap,sloped]{low} (root-neut-zero-low);
	\draw[->] (root-neut-zero) -- node [sloped]{high} (root-neut-zero-high);
	\draw[->] (root-neut-pos) -- node [swap,sloped]{low} (root-neut-pos-low);
	\draw[->] (root-neut-pos) -- node [sloped]{high} (root-neut-pos-high);
	\draw[->] (root-Nino-zero) -- node [swap,sloped]{low} (root-Nino-zero-low);
	\draw[->] (root-Nino-zero) -- node [sloped]{high} (root-Nino-zero-high);
	\draw[->] (root-Nino-pos) -- node [swap,sloped]{low} (root-Nino-pos-low);
	\draw[->] (root-Nino-pos) -- node [sloped]{high} (root-Nino-pos-high);
\end{tikzpicture}
    }
        \caption{Staged event tree from \cite{leonelli23a} illustrating the relationships among ENSO (El Niño Southern Oscillation), IOD (Indian Ocean Dipole), and AU (Australia precipitation), highlighting the estimated conditional probabilities for high AU in the three recovered stages. Nodes in the same stage are depicted with the same color, indicating shared conditional probabilities and context-specific independencies among the variables.}
    \label{fig:enso}
\end{figure}

As an example, we report in Fig.~\ref{fig:enso} the staged event tree learned by \cite{leonelli23a} on the effect of El Niño Southern Oscillation (ENSO) on Australian precipitation (AU) during spring, and the possible mediation of the Indian Ocean Dipole (IOD) as studied previously in \cite{QuantifyingCausalPathwaysofTeleconnections}.
The staged event tree, learned from observations, reveals a clear structure in the conditional probability tables: AU is independent of IOD, conditionally on ENSO being in the extreme phases of la Niña or el Niño and IOD is partially independent of ENSO ($P(\text{IOD}|\text{ENSO} = \text{Niño}) = P(\text{IOD}|\text{ENSO} = \text{neut})$.
The graphical representation in Fig.~\ref{fig:enso} is more interpretable than the DAG counterpart considered in \cite{QuantifyingCausalPathwaysofTeleconnections}, where similar conclusions were drawn by manually inspecting contingency tables. The staged event tree clearly highlights context-specific independencies, providing a more nuanced understanding of the causal relationships.

Although, from a modeling perspective, $\bm{X}$-compatible staged event trees are typically defined on a complete and symmetric event tree, in many real-world applications, certain scenarios within the tree may be physically or logically implausible, leading to unobserved combinations of variables (e.g., the El Niño and negative IOD phase in Fig.~\ref{fig:enso}). These \emph{unobserved} scenarios can be omitted from both the visual representation of the tree and the subsequent statistical analysis and learning algorithms. This allows $\bm X$-compatible staged event trees to effectively represent asymmetric sample spaces. In contrast, DAGs inherently obscure such asymmetries, as they only encode conditional independence relationships and do not explicitly represent unobserved scenarios. Such asymmetries can only be detected by closely examining the conditional probability tables.

\section{Casual Inference with Staged Event Trees}\label{sec:causalinference}

We consider here the task of estimating the ATE or its conditional version (CATE) for a binary treatment variable ($R$)  and a binary outcome ($Y$) when a set of additional categorical covariates ($\bm Z$) is available. While there are no methodological or computational barriers to extending this approach to more general categorical settings, we focus on this commonly studied case for clarity of exposition and ease of visualization.

We assume we only have access to observational data. Under the standard assumptions of consistency, positivity, and conditional exchangeability \cite{hernancausal}, the causal effect of $R$ on $Y$ is identifiable, and we can estimate it using statistical procedures. From a DAG perspective \cite{pearl2009causality}, conditional exchangeability is equivalent to assuming that the covariates $\mathbf{Z}$ are sufficient to \emph{control} (block) all paths from $R$ to $Y$ except the direct link $R \rightarrow Y$.

In this setting, different methods differ in the specific way they control for the covariates $\mathbf{Z}$. The two straightforward approaches are: (i) modeling the regression function or the conditional probability of $Y$ with respect to both $R$ and $\bm{Z}$; and (ii) modeling the propensity score, that is, the probability of treatment $P(R\,|\,\mathbf{Z})$. The regression-based approach (i) can be implemented with both parametric regression models (e.g., generalized linear models) or via standardization in a non-parametric fashion. Estimating the propensity score (ii) leads to inverse probability weighting (IPW), a generalization of the classical Horvitz--Thompson estimator of the mean \cite{horvitz1952generalization}, propensity score matching \cite{rosenbaum1983central}, or propensity score stratification and standardization via propensity score subclassification~\cite{rosenbaum1983central} among others. Moreover, the two approaches can be combined to obtain doubly robust estimators \cite{cao2009improving}, such as augmented IPW or double machine learning (DML) methods \cite{chernozhukov2018double}.

We refer to \cite{hernancausal} for a comprehensive introduction to causal inference methodologies from observational data. We simply point out that all the proposed approaches in this setting exploit the assumptions to achieve identifiability of the causal parameter of interest, and their variations are based on different functional assumptions (e.g., linearity) and on different finite-sample properties of the resulting estimators.

Staged event trees provide a flexible and expressive model class that is particularly well-suited for causal inference tasks. To illustrate how causal effect estimation can be based on a (causal) staged event tree, we present a detailed example from the literature. In \cite{aditi2019}  a hypothetical non-pharmacological intervention aimed at reducing the risk of falls among the elderly, who either live in the community or a communal establishment is explored. In this scenario, we imagine a stratified randomized controlled study being conducted to evaluate the intervention's effectiveness. Specifically, a proportion of elderly patients were assessed and categorized as either low or high risk. High-risk individuals could be referred to a clinic, and all referred patients, along with 50\% of non-referred high-risk individuals and 10\% of low-risk individuals, would receive treatment. Simulated data for this study was generated by \cite{aditi2019} and made available by \cite{walley2023cegpy}.

\begin{figure}[t!]
    \centering
    \resizebox{0.8\columnwidth}{!}{
        \input{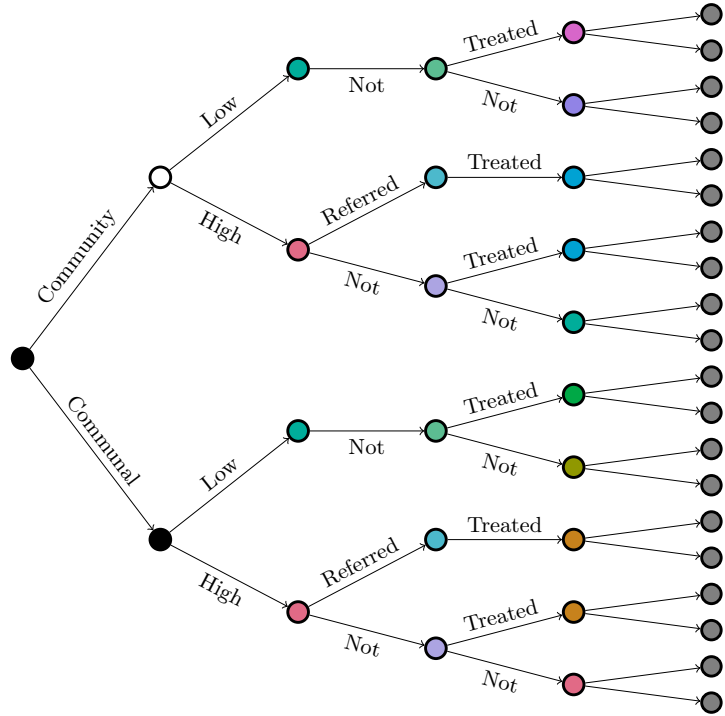}
    }
    \caption{Staged event tree for the Fall's intervention example. Variable ordering: Living situation (Communal/Community), Risk (Low/High), Referral (Referred/Not), Treatment (Treated/No), and Fall (Yes/No).}
    \label{fig:fall}
\end{figure}

In Fig.~\ref{fig:fall}, we depict the staged event tree aligned with the description and assumptions of this hypothetical study. Once the staged event tree is defined and probabilities are estimated from the data, causal effects can be readily derived. Specifically, the ATE can be computed by \textit{randomizing} the treatment variable and calculating conditional probabilities in the randomized staged event tree.
This procedure will be formally defined in the next section.
The staged event tree clearly illustrates a violation of the positivity assumption in the group of high-risk, referred patients, as all of them receive treatment. However, by assuming that the outcome is independent of referral status (conditional on covariates), we can still use observations from this group to estimate the treatment effect. Conditional treatment effects are directly observable from the transition probabilities for the outcome variable, and confidence intervals can be easily extracted when needed.

\subsection{Randomized staged event trees}

We assume a known \emph{causal order} of the variables, $\mathbf{X} = (\mathbf{Z}, R, Y)$, and the additionally the standard identifiability assumptions of consistency, positivity, and exchangeability. 

If the staged event tree $T$ is correctly specified, meaning that the data-generating process is contained within the model class $\mathcal{M}_T$, then, given sufficient observations, we can estimate the average treatment effect (ATE) by first estimating conditional probabilities from data and then randomizing $T$ on the treatment variable $R$. 
In details we define the \emph{randomized staged event tree} as follow. 

\begin{definition}[Randomized staged event tree]

Given a $\mathbf{X} = (\mathbf{Z}, R, Y)$ compatible staged even tree $T=(V, E, \eta)$ we define the associated randomized staged event tree as $T_r = (V, E, \eta_r)$ where 
$\eta_r(v, \mathbf{x}_{[i]}) = x_i$ if $X_i = R$ is the tratement variable\footnote{Meaning the associated $k_r$ function is a constant naive function for the treatment variable.} and $\eta_r(v, \mathbf{x}_{[i]}) = \eta(v, \mathbf{x}_{[i]})$ otherwise.
\end{definition}

This approach is equivalent to standardization with respect to the levels of $\mathbf{Z}$~\cite{hernancausal} also known as adjusting through subclassification \cite{rosenbaum1983central}. Under the stated assumptions, this leads to a consistent estimator for the ATE which follows from the following result which follow directly from the 
definition of staged tree causal model.

\begin{lemma}
    Let $P$ be a joint probability distribution
    over $\mathbf{X} = (\mathbf{Z}, R, Y)$. 
    Then for all values $r_0$ of the treatment variable, $$P(Z|\operatorname{do}(R = r_0)) = P_r(Z|R = r_0),$$ where 
    $P_r$ is any probability distribution in $\mathcal{M}_{T_r}$ such that 
    $$P_r(X_i|\mathbf{X}_{[i-1]}) = P(X_i|\mathbf{X}_{[i-1]}) \text{ for all } X_i \neq R.$$
\end{lemma}

When the staged event tree structure of the true data-generating model is unknown, we propose to combine one of the available stage structure learning routines with the randomized procedure described previously. The resulting estimators, while consistent under the assumption that the used stage learning routine retrieves the correct structure, are unlikely to be theoretically efficient and do not allow for a straightforward valid statistical inference. Nevertheless, similar approaches combining structural learning and effect estimation have been employed in the literature \cite{castelletti2021structural,maathuis2009estimating}. Moreover, the proposed strategy is similar to other approaches using ML algorithms such as LASSO regression \cite{shortreed2017outcome,tibshirani1996regression}, which employ variable or model selection techniques to improve the statistical properties of the final causal effect estimators. On this last point, we can hypothesize that by performing model selection and thus restricting the number of parameters in the final model, we can obtain ATE estimators with reduced variance compared to complete stratification, which is equivalent to considering a randomized staged event tree where every node is in an individual stage.


\subsection{PS-Stratified Staged Event Trees}

Causal effects estimated by randomizing the staged event tree correspond to adjustment via standardization
with respect to the considered covariates. 
Such a method employs only the information encoded in the stages structure of the outcome variable (the last level of the staged event tree).

Alternatively, as we summarized previously, ATEs can be estimated by employing the information of the treatment assignment, that is by modeling the propensity scores. 
In a staged event tree, this is naturally encoded in the stages' structure of the treatment variable.
By modifying accordingly the staged event tree is thus possible to implement a simple form of propensity score stratification~\cite{rosenbaum1983central}.

In particular, we define the \emph{ps-stratified} staged event tree as the tree where the stages of the outcome variable are generated from the treatment stages as follows: 
if two situations are in the same treatment stage then we impose the same stages for the outcome in the corresponding sub-trees.
Specifically, 
\begin{definition}[The ps-stratified staged event tree]

    Given an $\mathbf{X} = (\mathbf{Z}, R, Y)$ compatible staged even tree $T=(V, E, \eta)$ we define the associated ps-stratified staged event tree as $T_{ps} = (V, E, \eta_{ps})$ where 
$\eta_{ps}(\mathbf{x}_{[i-1]}, \mathbf{x}_{[i]}) = (\eta(\mathbf{x}_{[i-2]}, \mathbf{x}_{[i-1]}), x_i)$ if $X_i = Y$ is the outcome variable and $\eta_{ps}(v, \mathbf{x}_{[i]}) = \eta(v, \mathbf{x}_{[i]})$ otherwise.

\end{definition}

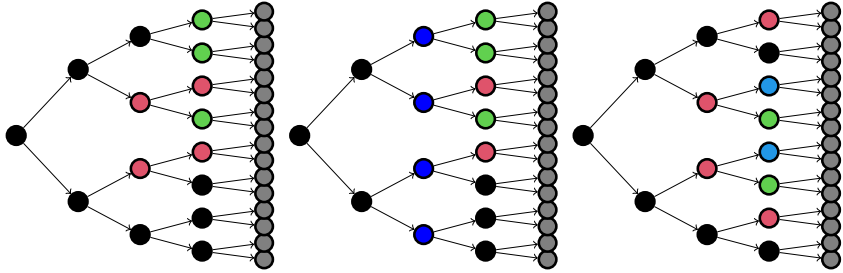
\begin{figure}
    \resizebox{0.3\columnwidth}{!}{\begin{tikzpicture}[auto, scale=4,
	X1_NA/.style={circle,inner sep=1mm,minimum size=0.3cm,draw,black,very thick,fill={rgb,255:red,0; green,0; blue,0},text=black},
	X2_1/.style={circle,inner sep=1mm,minimum size=0.3cm,draw,black,very thick,fill={rgb,255:red,0; green,0; blue,0},text=black},
	A_1/.style={circle,inner sep=1mm,minimum size=0.3cm,draw,black,very thick,fill={rgb,255:red,0; green,0; blue,0},text=black},
	A_2/.style={circle,inner sep=1mm,minimum size=0.3cm,draw,black,very thick,fill={rgb,255:red,223; green,83; blue,107},text=black},
	Y_1/.style={circle,inner sep=1mm,minimum size=0.3cm,draw,black,very thick,fill={rgb,255:red,0; green,0; blue,0},text=black},
	Y_2/.style={circle,inner sep=1mm,minimum size=0.3cm,draw,black,very thick,fill={rgb,255:red,223; green,83; blue,107},text=black},
	Y_3/.style={circle,inner sep=1mm,minimum size=0.3cm,draw,black,very thick,fill={rgb,255:red,97; green,208; blue,79},text=black},
	 leaf/.style={circle,inner sep=1mm,minimum size=0.2cm,draw,very thick,black,fill=gray,text=black}]

	\node [X1_NA] (root) at (0.000000, 0.500000)	{};
	\node [X2_1] (root-a) at (0.250000, 0.233333)	{};
	\node [X2_1] (root-b) at (0.250000, 0.766667)	{};
	\node [A_1] (root-a-0) at (0.500000, 0.100000)	{};
	\node [A_2] (root-a-1) at (0.500000, 0.366667)	{};
	\node [A_2] (root-b-0) at (0.500000, 0.633333)	{};
	\node [A_1] (root-b-1) at (0.500000, 0.900000)	{};
	\node [Y_1] (root-a-0-0) at (0.750000, 0.033333)	{};
	\node [Y_1] (root-a-0-1) at (0.750000, 0.166667)	{};
	\node [Y_1] (root-a-1-0) at (0.750000, 0.300000)	{};
	\node [Y_2] (root-a-1-1) at (0.750000, 0.433333)	{};
	\node [Y_3] (root-b-0-0) at (0.750000, 0.566667)	{};
	\node [Y_2] (root-b-0-1) at (0.750000, 0.700000)	{};
	\node [Y_3] (root-b-1-0) at (0.750000, 0.833333)	{};
	\node [Y_3] (root-b-1-1) at (0.750000, 0.966667)	{};
	\node [leaf] (root-a-0-0-+1) at (1.000000, 0.000000)	{};
	\node [leaf] (root-a-0-0--1) at (1.000000, 0.066667)	{};
	\node [leaf] (root-a-0-1-+1) at (1.000000, 0.133333)	{};
	\node [leaf] (root-a-0-1--1) at (1.000000, 0.200000)	{};
	\node [leaf] (root-a-1-0-+1) at (1.000000, 0.266667)	{};
	\node [leaf] (root-a-1-0--1) at (1.000000, 0.333333)	{};
	\node [leaf] (root-a-1-1-+1) at (1.000000, 0.400000)	{};
	\node [leaf] (root-a-1-1--1) at (1.000000, 0.466667)	{};
	\node [leaf] (root-b-0-0-+1) at (1.000000, 0.533333)	{};
	\node [leaf] (root-b-0-0--1) at (1.000000, 0.600000)	{};
	\node [leaf] (root-b-0-1-+1) at (1.000000, 0.666667)	{};
	\node [leaf] (root-b-0-1--1) at (1.000000, 0.733333)	{};
	\node [leaf] (root-b-1-0-+1) at (1.000000, 0.800000)	{};
	\node [leaf] (root-b-1-0--1) at (1.000000, 0.866667)	{};
	\node [leaf] (root-b-1-1-+1) at (1.000000, 0.933333)	{};
	\node [leaf] (root-b-1-1--1) at (1.000000, 1.000000)	{};

	\draw[->] (root) -- node [sloped]{} (root-a);
	\draw[->] (root) -- node [sloped]{} (root-b);
	\draw[->] (root-a) -- node [sloped]{} (root-a-0);
	\draw[->] (root-a) -- node [sloped]{} (root-a-1);
	\draw[->] (root-b) -- node [sloped]{} (root-b-0);
	\draw[->] (root-b) -- node [sloped]{} (root-b-1);
	\draw[->] (root-a-0) -- node [sloped]{} (root-a-0-0);
	\draw[->] (root-a-0) -- node [sloped]{} (root-a-0-1);
	\draw[->] (root-a-1) -- node [sloped]{} (root-a-1-0);
	\draw[->] (root-a-1) -- node [sloped]{} (root-a-1-1);
	\draw[->] (root-b-0) -- node [sloped]{} (root-b-0-0);
	\draw[->] (root-b-0) -- node [sloped]{} (root-b-0-1);
	\draw[->] (root-b-1) -- node [sloped]{} (root-b-1-0);
	\draw[->] (root-b-1) -- node [sloped]{} (root-b-1-1);
	\draw[->] (root-a-0-0) -- node [sloped]{} (root-a-0-0-+1);
	\draw[->] (root-a-0-0) -- node [sloped]{} (root-a-0-0--1);
	\draw[->] (root-a-0-1) -- node [sloped]{} (root-a-0-1-+1);
	\draw[->] (root-a-0-1) -- node [sloped]{} (root-a-0-1--1);
	\draw[->] (root-a-1-0) -- node [sloped]{} (root-a-1-0-+1);
	\draw[->] (root-a-1-0) -- node [sloped]{} (root-a-1-0--1);
	\draw[->] (root-a-1-1) -- node [sloped]{} (root-a-1-1-+1);
	\draw[->] (root-a-1-1) -- node [sloped]{} (root-a-1-1--1);
	\draw[->] (root-b-0-0) -- node [sloped]{} (root-b-0-0-+1);
	\draw[->] (root-b-0-0) -- node [sloped]{} (root-b-0-0--1);
	\draw[->] (root-b-0-1) -- node [sloped]{} (root-b-0-1-+1);
	\draw[->] (root-b-0-1) -- node [sloped]{} (root-b-0-1--1);
	\draw[->] (root-b-1-0) -- node [sloped]{} (root-b-1-0-+1);
	\draw[->] (root-b-1-0) -- node [sloped]{} (root-b-1-0--1);
	\draw[->] (root-b-1-1) -- node [sloped]{} (root-b-1-1-+1);
	\draw[->] (root-b-1-1) -- node [sloped]{} (root-b-1-1--1);
\end{tikzpicture}}
    \resizebox{0.3\columnwidth}{!}{\begin{tikzpicture}[auto, scale=4,
	X1_NA/.style={circle,inner sep=1mm,minimum size=0.3cm,draw,black,very thick,fill={rgb,255:red,0; green,0; blue,0},text=black},
	X2_1/.style={circle,inner sep=1mm,minimum size=0.3cm,draw,black,very thick,fill={rgb,255:red,0; green,0; blue,0},text=black},
	A_randomized/.style={circle,inner sep=1mm,minimum size=0.3cm,draw,black,very thick,fill={rgb,255:red,0; green,0; blue,255},text=black},
	Y_1/.style={circle,inner sep=1mm,minimum size=0.3cm,draw,black,very thick,fill={rgb,255:red,0; green,0; blue,0},text=black},
	Y_2/.style={circle,inner sep=1mm,minimum size=0.3cm,draw,black,very thick,fill={rgb,255:red,223; green,83; blue,107},text=black},
	Y_3/.style={circle,inner sep=1mm,minimum size=0.3cm,draw,black,very thick,fill={rgb,255:red,97; green,208; blue,79},text=black},
	 leaf/.style={circle,inner sep=1mm,minimum size=0.2cm,draw,very thick,black,fill=gray,text=black}]

	\node [X1_NA] (root) at (0.000000, 0.500000)	{};
	\node [X2_1] (root-a) at (0.250000, 0.233333)	{};
	\node [X2_1] (root-b) at (0.250000, 0.766667)	{};
	\node [A_randomized] (root-a-0) at (0.500000, 0.100000)	{};
	\node [A_randomized] (root-a-1) at (0.500000, 0.366667)	{};
	\node [A_randomized] (root-b-0) at (0.500000, 0.633333)	{};
	\node [A_randomized] (root-b-1) at (0.500000, 0.900000)	{};
	\node [Y_1] (root-a-0-0) at (0.750000, 0.033333)	{};
	\node [Y_1] (root-a-0-1) at (0.750000, 0.166667)	{};
	\node [Y_1] (root-a-1-0) at (0.750000, 0.300000)	{};
	\node [Y_2] (root-a-1-1) at (0.750000, 0.433333)	{};
	\node [Y_3] (root-b-0-0) at (0.750000, 0.566667)	{};
	\node [Y_2] (root-b-0-1) at (0.750000, 0.700000)	{};
	\node [Y_3] (root-b-1-0) at (0.750000, 0.833333)	{};
	\node [Y_3] (root-b-1-1) at (0.750000, 0.966667)	{};
	\node [leaf] (root-a-0-0-+1) at (1.000000, 0.000000)	{};
	\node [leaf] (root-a-0-0--1) at (1.000000, 0.066667)	{};
	\node [leaf] (root-a-0-1-+1) at (1.000000, 0.133333)	{};
	\node [leaf] (root-a-0-1--1) at (1.000000, 0.200000)	{};
	\node [leaf] (root-a-1-0-+1) at (1.000000, 0.266667)	{};
	\node [leaf] (root-a-1-0--1) at (1.000000, 0.333333)	{};
	\node [leaf] (root-a-1-1-+1) at (1.000000, 0.400000)	{};
	\node [leaf] (root-a-1-1--1) at (1.000000, 0.466667)	{};
	\node [leaf] (root-b-0-0-+1) at (1.000000, 0.533333)	{};
	\node [leaf] (root-b-0-0--1) at (1.000000, 0.600000)	{};
	\node [leaf] (root-b-0-1-+1) at (1.000000, 0.666667)	{};
	\node [leaf] (root-b-0-1--1) at (1.000000, 0.733333)	{};
	\node [leaf] (root-b-1-0-+1) at (1.000000, 0.800000)	{};
	\node [leaf] (root-b-1-0--1) at (1.000000, 0.866667)	{};
	\node [leaf] (root-b-1-1-+1) at (1.000000, 0.933333)	{};
	\node [leaf] (root-b-1-1--1) at (1.000000, 1.000000)	{};

	\draw[->] (root) -- node [sloped]{} (root-a);
	\draw[->] (root) -- node [sloped]{} (root-b);
	\draw[->] (root-a) -- node [sloped]{} (root-a-0);
	\draw[->] (root-a) -- node [sloped]{} (root-a-1);
	\draw[->] (root-b) -- node [sloped]{} (root-b-0);
	\draw[->] (root-b) -- node [sloped]{} (root-b-1);
	\draw[->] (root-a-0) -- node [sloped]{} (root-a-0-0);
	\draw[->] (root-a-0) -- node [sloped]{} (root-a-0-1);
	\draw[->] (root-a-1) -- node [sloped]{} (root-a-1-0);
	\draw[->] (root-a-1) -- node [sloped]{} (root-a-1-1);
	\draw[->] (root-b-0) -- node [sloped]{} (root-b-0-0);
	\draw[->] (root-b-0) -- node [sloped]{} (root-b-0-1);
	\draw[->] (root-b-1) -- node [sloped]{} (root-b-1-0);
	\draw[->] (root-b-1) -- node [sloped]{} (root-b-1-1);
	\draw[->] (root-a-0-0) -- node [sloped]{} (root-a-0-0-+1);
	\draw[->] (root-a-0-0) -- node [sloped]{} (root-a-0-0--1);
	\draw[->] (root-a-0-1) -- node [sloped]{} (root-a-0-1-+1);
	\draw[->] (root-a-0-1) -- node [sloped]{} (root-a-0-1--1);
	\draw[->] (root-a-1-0) -- node [sloped]{} (root-a-1-0-+1);
	\draw[->] (root-a-1-0) -- node [sloped]{} (root-a-1-0--1);
	\draw[->] (root-a-1-1) -- node [sloped]{} (root-a-1-1-+1);
	\draw[->] (root-a-1-1) -- node [sloped]{} (root-a-1-1--1);
	\draw[->] (root-b-0-0) -- node [sloped]{} (root-b-0-0-+1);
	\draw[->] (root-b-0-0) -- node [sloped]{} (root-b-0-0--1);
	\draw[->] (root-b-0-1) -- node [sloped]{} (root-b-0-1-+1);
	\draw[->] (root-b-0-1) -- node [sloped]{} (root-b-0-1--1);
	\draw[->] (root-b-1-0) -- node [sloped]{} (root-b-1-0-+1);
	\draw[->] (root-b-1-0) -- node [sloped]{} (root-b-1-0--1);
	\draw[->] (root-b-1-1) -- node [sloped]{} (root-b-1-1-+1);
	\draw[->] (root-b-1-1) -- node [sloped]{} (root-b-1-1--1);
\end{tikzpicture}}
    \resizebox{0.3\columnwidth}{!}{\begin{tikzpicture}[auto, scale=4,
	X1_NA/.style={circle,inner sep=1mm,minimum size=0.3cm,draw,black,very thick,fill={rgb,255:red,0; green,0; blue,0},text=black},
	X2_1/.style={circle,inner sep=1mm,minimum size=0.3cm,draw,black,very thick,fill={rgb,255:red,0; green,0; blue,0},text=black},
	A_1/.style={circle,inner sep=1mm,minimum size=0.3cm,draw,black,very thick,fill={rgb,255:red,0; green,0; blue,0},text=black},
	A_2/.style={circle,inner sep=1mm,minimum size=0.3cm,draw,black,very thick,fill={rgb,255:red,223; green,83; blue,107},text=black},
	Y_11/.style={circle,inner sep=1mm,minimum size=0.3cm,draw,black,very thick,fill={rgb,255:red,0; green,0; blue,0},text=black},
	Y_12/.style={circle,inner sep=1mm,minimum size=0.3cm,draw,black,very thick,fill={rgb,255:red,223; green,83; blue,107},text=black},
	Y_21/.style={circle,inner sep=1mm,minimum size=0.3cm,draw,black,very thick,fill={rgb,255:red,97; green,208; blue,79},text=black},
	Y_22/.style={circle,inner sep=1mm,minimum size=0.3cm,draw,black,very thick,fill={rgb,255:red,34; green,151; blue,230},text=black},
	 leaf/.style={circle,inner sep=1mm,minimum size=0.2cm,draw,very thick,black,fill=gray,text=black}]

	\node [X1_NA] (root) at (0.000000, 0.500000)	{};
	\node [X2_1] (root-a) at (0.250000, 0.233333)	{};
	\node [X2_1] (root-b) at (0.250000, 0.766667)	{};
	\node [A_1] (root-a-0) at (0.500000, 0.100000)	{};
	\node [A_2] (root-a-1) at (0.500000, 0.366667)	{};
	\node [A_2] (root-b-0) at (0.500000, 0.633333)	{};
	\node [A_1] (root-b-1) at (0.500000, 0.900000)	{};
	\node [Y_11] (root-a-0-0) at (0.750000, 0.033333)	{};
	\node [Y_12] (root-a-0-1) at (0.750000, 0.166667)	{};
	\node [Y_21] (root-a-1-0) at (0.750000, 0.300000)	{};
	\node [Y_22] (root-a-1-1) at (0.750000, 0.433333)	{};
	\node [Y_21] (root-b-0-0) at (0.750000, 0.566667)	{};
	\node [Y_22] (root-b-0-1) at (0.750000, 0.700000)	{};
	\node [Y_11] (root-b-1-0) at (0.750000, 0.833333)	{};
	\node [Y_12] (root-b-1-1) at (0.750000, 0.966667)	{};
	\node [leaf] (root-a-0-0-+1) at (1.000000, 0.000000)	{};
	\node [leaf] (root-a-0-0--1) at (1.000000, 0.066667)	{};
	\node [leaf] (root-a-0-1-+1) at (1.000000, 0.133333)	{};
	\node [leaf] (root-a-0-1--1) at (1.000000, 0.200000)	{};
	\node [leaf] (root-a-1-0-+1) at (1.000000, 0.266667)	{};
	\node [leaf] (root-a-1-0--1) at (1.000000, 0.333333)	{};
	\node [leaf] (root-a-1-1-+1) at (1.000000, 0.400000)	{};
	\node [leaf] (root-a-1-1--1) at (1.000000, 0.466667)	{};
	\node [leaf] (root-b-0-0-+1) at (1.000000, 0.533333)	{};
	\node [leaf] (root-b-0-0--1) at (1.000000, 0.600000)	{};
	\node [leaf] (root-b-0-1-+1) at (1.000000, 0.666667)	{};
	\node [leaf] (root-b-0-1--1) at (1.000000, 0.733333)	{};
	\node [leaf] (root-b-1-0-+1) at (1.000000, 0.800000)	{};
	\node [leaf] (root-b-1-0--1) at (1.000000, 0.866667)	{};
	\node [leaf] (root-b-1-1-+1) at (1.000000, 0.933333)	{};
	\node [leaf] (root-b-1-1--1) at (1.000000, 1.000000)	{};

	\draw[->] (root) -- node [sloped]{} (root-a);
	\draw[->] (root) -- node [sloped]{} (root-b);
	\draw[->] (root-a) -- node [sloped]{} (root-a-0);
	\draw[->] (root-a) -- node [sloped]{} (root-a-1);
	\draw[->] (root-b) -- node [sloped]{} (root-b-0);
	\draw[->] (root-b) -- node [sloped]{} (root-b-1);
	\draw[->] (root-a-0) -- node [sloped]{} (root-a-0-0);
	\draw[->] (root-a-0) -- node [sloped]{} (root-a-0-1);
	\draw[->] (root-a-1) -- node [sloped]{} (root-a-1-0);
	\draw[->] (root-a-1) -- node [sloped]{} (root-a-1-1);
	\draw[->] (root-b-0) -- node [sloped]{} (root-b-0-0);
	\draw[->] (root-b-0) -- node [sloped]{} (root-b-0-1);
	\draw[->] (root-b-1) -- node [sloped]{} (root-b-1-0);
	\draw[->] (root-b-1) -- node [sloped]{} (root-b-1-1);
	\draw[->] (root-a-0-0) -- node [sloped]{} (root-a-0-0-+1);
	\draw[->] (root-a-0-0) -- node [sloped]{} (root-a-0-0--1);
	\draw[->] (root-a-0-1) -- node [sloped]{} (root-a-0-1-+1);
	\draw[->] (root-a-0-1) -- node [sloped]{} (root-a-0-1--1);
	\draw[->] (root-a-1-0) -- node [sloped]{} (root-a-1-0-+1);
	\draw[->] (root-a-1-0) -- node [sloped]{} (root-a-1-0--1);
	\draw[->] (root-a-1-1) -- node [sloped]{} (root-a-1-1-+1);
	\draw[->] (root-a-1-1) -- node [sloped]{} (root-a-1-1--1);
	\draw[->] (root-b-0-0) -- node [sloped]{} (root-b-0-0-+1);
	\draw[->] (root-b-0-0) -- node [sloped]{} (root-b-0-0--1);
	\draw[->] (root-b-0-1) -- node [sloped]{} (root-b-0-1-+1);
	\draw[->] (root-b-0-1) -- node [sloped]{} (root-b-0-1--1);
	\draw[->] (root-b-1-0) -- node [sloped]{} (root-b-1-0-+1);
	\draw[->] (root-b-1-0) -- node [sloped]{} (root-b-1-0--1);
	\draw[->] (root-b-1-1) -- node [sloped]{} (root-b-1-1-+1);
	\draw[->] (root-b-1-1) -- node [sloped]{} (root-b-1-1--1);
\end{tikzpicture}}
    \caption{Original (left), randomized  (center) and ps-stratified  (right) staged event trees.}
    \label{fig:transform}
\end{figure}

An example of the randomized and ps-stratified transformations is depicted in Fig.~\ref{fig:transform}, together with the original staged event tree.

\subsection{Uncertainty}

A central goal of causal inference is to provide valid uncertainty for the 
point estimators proposed. This is because one task of causal inference is, for instance, testing \emph{if a causal effect is effectively present}.
Thus, appropriate ATE and CATE estimators should be accompanied by standard error estimates, (asymptotic) normality results, or valid confidence intervals. 
The proposed approach, based on learning a parsimonious model (for the outcome regression or the propensity score) and sub-sequentially obtaining ATE and CATE estimates, suffers inevitably from post-selection inference problems \cite{berk2013valid}. 
A possible solution is the use of bootstrap to generate valid confidence intervals, an approach already explored in the staged event tree literature for learning robust models~\cite{LEONELLI2024102030}.
We show in Section~\ref{sec:experiments2} how through bootstrap we can easily obtain uncertainty quantification. 

\section{Simulation Experiments}\label{sec:experiments}

We exploit the staged event trees' implementation in the \texttt{stagedtrees} R package~\cite{carli2022}. 
We define staged tree estimators by learning the structure of the stages with a fixed order of variables $\mathbf{X} = (\mathbf{Z}, R, Y)$ and then randomizing the treatment variable $R$ and computing $P(Y| \operatorname{do}(R \sim \text{Bernulli} (0.5)))$ where the operator $\operatorname{do}(R \sim \text{Bernulli} (0.5))$ denotes the intervention resulting in the randomization of $R$ in the learned staged tree. 
Alternatively, we consider the ATE estimated via the ps-stratified staged event tree by the weighted average of the strata defined by the stages of the treatment variable. 

We couple the above two estimators with the following algorithms for stage structure learning:
\begin{itemize}
    \item \texttt{BHC} - a backward hill climbing approach similar to the classical AHC algorithm~\cite{freeman2011bayesian} but optimizing the BIC score (which has desiderable properties, see~\cite{gorgen2022curved}). 
    \item \texttt{hclust} - a search of the best clustering in stages, which combines a hierarchical clustering of conditional probabilities using total variation distance and 
    a search of the cut of the obtained dendrograms optimizing the BIC score. 
\end{itemize}

We compare the two obtained ATE estimators with classical ones: 
\begin{itemize}
    \item \texttt{AIPW} - the doubly-robust estimator~\cite{bang2005doubly} implemented in the R package \texttt{targeted} and using generalized linear models for the logistic regressions with only main effects (no interactions).
    \item \texttt{q.model} - the regression-based approach obtained with the same function as the \texttt{AIPW} but specifying a constant regression model for the propensity.
    \item \texttt{BN\_tabu} and \texttt{BN\_pc} -
    structural learning of a DAG with the \texttt{bnlearn} R package (\texttt{tabu} and \texttt{pc.stable} functions), randomizing the treatment variable;
    \item \texttt{full} - complete stratification obtained by fitting a complete staged event tree and randomizing the treatment $R$.
    \item \texttt{oracle} - assuming the knowledge of the true data generating model (the stages structure and only fitting probabilities from observations). 
\end{itemize}

We compare the above methods on simulation experiments where we sample data from random staged event trees obtained from full event trees (\texttt{sevt}) or DAGs (\texttt{dag}) with 
different probabilities ($0$, $0.5$ and $0.8$) of random stages joining. Conditional probabilities for the data-generating trees are sampled either with exponential or uniform distributions (\texttt{exp} or \texttt{unif}) and normalized to sum up to one.
We consider $6$ binary covariates and sample sizes $N=100, 500, 1000, 10000$. We consider 20 random repetitions for each setting and report median values across repetitions. 
We evaluate the ATE estimation by considering the absolute errors concerning the true ATE obtained by randomizing the true data-generating trees.

\begin{figure*}
    \centering
    \includegraphics[width = \linewidth]{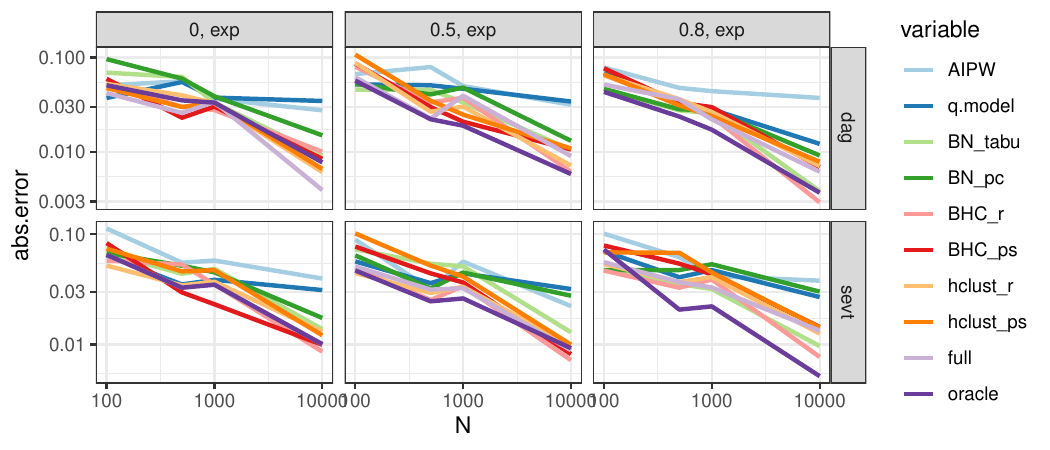}
    \caption{Median absolute error between true and estimated ATE for the considered estimators (case with $p=8$). In different columns, different data generating settings.}
    \label{fig:res8}
\end{figure*}

In Fig.~\ref{fig:res8}, we report the median absolute error of the different estimators as a function of the sample size.
As we can see, all methods converge towards the true value. Not surprisingly, the \texttt{oracle} estimator obtains the best performances across all settings.
The \texttt{full}, \texttt{hclust}, and \texttt{BHC} estimators have comparable results, showing that learning staged event trees is a valid procedure for ATE estimation. 
DAG-based estimators perform comparably to the staged event tree based methods while 
the remaining estimators have more diverse performances across sample sizes and data-generating settings; this is to be expected as \texttt{q.model} and \texttt{AIPW} do not use a correctly specified model class and thus, their theoretical guarantees do not hold.


\section{Data Analyses}\label{sec:experiments2}

We now showcase the potential of staged event trees in real-world  and complex 
causal inference problems.
With this goal, we consider two classical datasets in the causal inference literature, \emph{the electronic fetal monitoring and cesarean section} dataset from \cite{neutra1980effect} and the \emph{right heart catheterization} dataset~\cite{connors1996effectiveness}. 
Both datasets are available through the \texttt{ATbounds} R package~\cite{atbounds}.

\subsection{Effect of Fetal Monitoring on Cesarean Section Rates}

\begin{figure*}
    \centering
    \includegraphics[width=\textwidth]{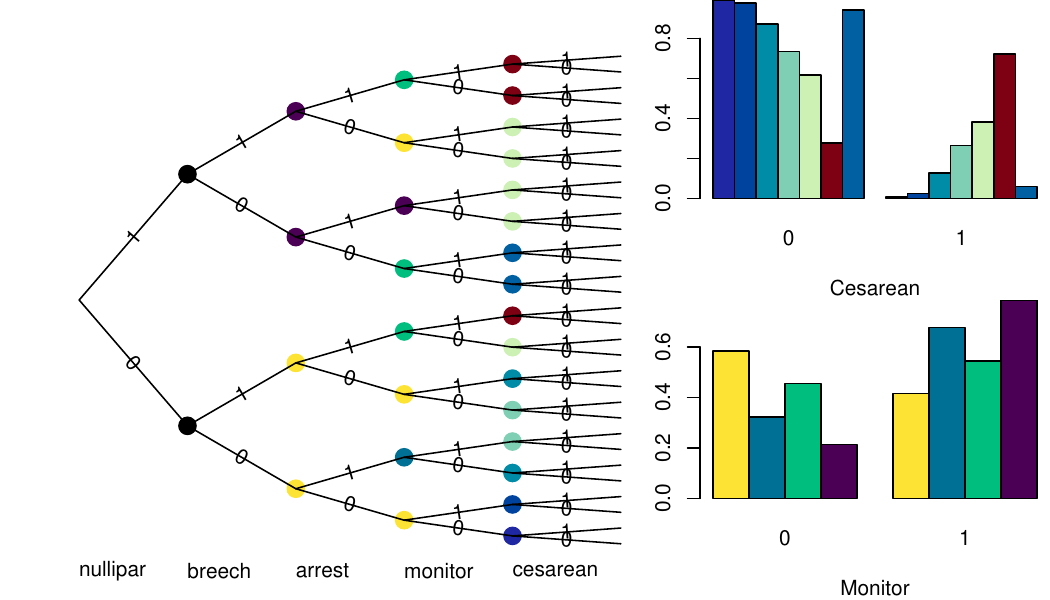}
    \caption{Staged event tree obtained with the \texttt{stages\_bhc} method for the EFM dataset, together with estimated conditional probabilities for treatment (Monitor) and outcome (Cesarean) in the learned stages.}
    \label{fig:EFM}
\end{figure*}

We replicate the analysis of the \emph{the electronic fetal monitoring (EFM) and cesarean section (CS)} dataset from \cite{neutra1980effect}. In particular, we model a staged event tree over all the variables (covariates, treatment, and outcome). We employ the 
hierarchical clustering method, as explained in the previous section.  The resulting tree with the estimated conditional probabilities is shown in Fig.~\ref{fig:EFM}.

We obtain comparable results to the literature~\cite{neutra1980effect,richardson2017modeling}, in 
particular we obtain opposite effects for the two subgroups: no-nullipar, breech, no-arrest and no-nullipar, no-breach, arrest. 

Differently from the previous study~\cite{richardson2017modeling}, the staged tree modeling estimates no-effect for the nullipar group. 
This difference could be explained by the fact that \cite{richardson2017modeling} uses additional temporal information and estimates time-varying effect curves while we 
aggregate all the data in the study for simplicity of exposition. 
Additionally, we observe that the model depicted in Fig.~\ref{fig:EFM} is highly interpretable, facilitating a storytelling approach to causal effect estimation. By explicitly representing the relationships between covariates and treatments, the staged tree model provides a more comprehensive and structured view of the problem, making it a valuable tool for practitioners.

Confidence intervals for the outcome probabilities given the treatment values across different strata can be obtained using standard inferential methods for categorical data (see~\cite{carli2022}). This allows for formal inference on the conditional average treatment effect (CATE). 

\subsection{Right Heart Catheterization}

In this section, we analyze a more complex dataset on the Right Heart Catheterization (RHC) procedure~\cite{connors1996effectiveness}. This problem is well-studied in the causal inference literature~\cite{li2009efficient}, as RHC is a widely used diagnostic tool that enables direct measurement of cardiac function. However, its effectiveness has been the subject of ongoing debate, and ethical concerns have prevented doctors from conducting a randomized trial to assess its impact.

We showcase here how the staged event tree approaches can deliver uncertainty estimates via bootstrap replicates.
In particular, for each replicate, we employ the hierarchical clustering method on empirical probabilities to learn the structure of the stages while selecting the number of stages by maximizing the BIC score. 
We then employ both the randomization and the propensity score stratification approaches to estimate the ATE.

\begin{figure}[h!]
    \centering
    \includegraphics[width=0.8\linewidth]{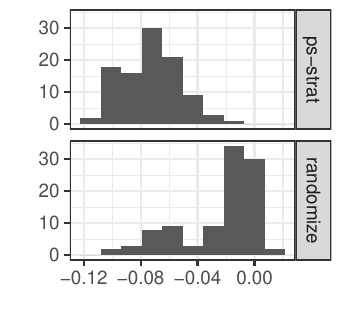}
    \caption{Distribution across the bootstrap replicates for the ps-stratified and randomized ATE estimates.}
    \label{fig:histRHC}
\end{figure}

Fig.~\ref{fig:histRHC} shows the distribution of the ATE estimates with the two proposed approaches. 
The ps-stratified staged event tree obtains an average (across bootstrap replicates) ATE equal to -0.0722 (95\% CI: $-0.107,-0.029$)  while the randomized staged event tree finds an average ATE estimate of $-0.0248$ (95\% CI: $-0.083, 0.0051$).
Both methods obtain an average negative effect, in line with some of the previous approaches~\cite{hirano2001estimation}. Moreover ps-stratified obtained an effect significantly (at level 0.05) different from zero.

A key advantage of staged event trees is their ability to naturally incorporate uncertainty quantification for causal estimands through resampling techniques such as the bootstrap. Unlike many standard causal inference methods, where deriving confidence intervals for treatment effects often requires complex variance estimation techniques, staged trees allow for a straightforward and computationally efficient approach. By leveraging their flexible structure, practitioners can readily obtain uncertainty measures for critical causal quantities, making staged event trees a valuable tool for robust causal inference.

\section{Conclusions}\label{sec:conclusions}


We demonstrated how staged event trees provide a powerful and interpretable approach to estimating treatment effects using observational data, extending previous work on probabilistic graphical modeling and causal discovery. By leveraging their ability to model complex, asymmetric dependencies, staged event trees enhance the precision and flexibility of causal inference while maintaining transparency. Additionally, this paper introduced routines that integrate the computation of causal estimators directly within the staged event tree framework, making it easier to embed causal inference techniques in this setting.

Through both simulation studies and real-world datasets, we illustrated the efficacy of staged event tree estimators in comparison to classical methods, highlighting their robust performance across various settings. The results show that staged event trees can capture intricate dependency structures that standard approaches often overlook, leading to more refined treatment effect estimates. In particular, we demonstrated how they can provide meaningful stratifications of the covariate space and deliver uncertainty quantification through resampling techniques. While this approach is not necessarily theoretically optimal, it offers a compelling alternative for practitioners seeking a balance between model flexibility and interpretability.

Future work may focus on refining staged event tree learning algorithms and expanding their application to broader domains, including Earth and Climate sciences, as well as Brain sciences. A natural next step is to explore the incorporation of time-varying estimators, particularly relevant for applications such as fetal monitoring, by leveraging recent developments in longitudinal staged event trees \cite{carter2025staged}. Another promising direction is the development of a fully Bayesian approach for model selection and estimation, similar to what has been done for DAGs \cite{castelletti2021structural}, which would not only refine staged tree construction but also provide automatic uncertainty quantification for causal estimates.

Ultimately, staged event trees represent a valuable tool in the expanding landscape of causal inference methodologies. By offering a balance of accuracy, efficiency, and ease of interpretation, they provide a practical and intuitive framework for estimating causal effects in complex real-world settings.



%
%
%
\bibliographystyle{splncs04}
\bibliography{biblio}
\end{document}